# Systems Biology: From the Cell to the Brain

**SITABHRA SINHA\*, T JESAN† and NIVEDITA CHATTERJEE+**
\* The Institute of Mathematical Sciences, CIT Campus, Taramani, Chennai 600113.
† Health Physics Division, Bhabha Atomic Research Centre, Kalpakkam 603201.
+Vision Research Foundation, Sankara Nethralaya, 18 College Road, Chennai 600006.

*Email\**: *sitabhra@imsc.res.in*

*Abstract*

*With the completion of human genome mapping, the focus of scientists seeking to explain the biological complexity of living systems is shifting from analyzing the individual components (such as a particular gene or biochemical reaction) to understanding the set of interactions amongst the large number of components that results in the different functions of the organism. To this end, the area of systems biology attempts to achieve a 'systems-level' description of biology by focussing on the network of interactions instead of the characteristics of its isolated parts. In this article, we briefly describe some of the emerging themes of research in 'network' biology, looking at dynamical processes occurring at the two different length scales of within the cell and between cells, viz., the intra-cellular signaling network and the nervous system. We show that focusing on the systems-level aspects of these problems allows one to observe surprising and illuminating common themes amongst them.*

**INTRODUCTION**

The complete mapping of human and other genomes has indicated that the remarkable complexity of living organisms is expressed by less than 30,000 protein-coding genes [1]. Thus, the observed complexity arises not so much from the relatively few components (in this case, genes), as from the large set of mutual interactions that they are capable of generating. In a similar fashion, the 302 neurons of the nematode *C. elegans* enables it to survive in the wild, much more successfully than complicated, state-of-the-art robots. It is not the number of neurons that is the crucial factor here, but rather their interactions and the resulting repertoire of dynamical responses that underlie the survival success of living organisms in a hostile (and often unpredictable) environment. The focus of research in biology is therefore gradually shifting towards understanding how interactions between components, be they genes, proteins, cells or organisms, add a qualitatively new layer of complexity to the biological world. This is the domain of *systems biology* which aims to understand organisms as an integrated whole of interacting genetic, protein and biochemical reaction networks, rather than focusing on the individual components in isolation [2-4]. While the term itself is of recent coinage [2], the field has had several antecedents, most notably, *cybernetics*, as pioneered by Norbert Wiener [5] and W Ross Ashby (who indeed can be considered to be one of the founding figures of systems neuroscience [6] along with Warren McCulloch [7]) and the *general systems theory* of Ludwig von Bertalanffy, which have inspired other fields in addition to biology.

The recent surge of interest in systems thinking in biology has been fuelled by the fortunate coincidence in the advent of high throughput experimental techniques (such as DNA and protein microarrays) allowing multiplex assays, along with the almost simultaneous development of affordable high-performance computing which has made possible automated analysis of huge volumes of experimental data and the simulation of very large complex systems. Another possible stimulant has been the parallel growth of the theory of *complex networks* (comprising many nodes that are connected by links arranged according to some non-trivial topology) from 1998 onwards, which has provided a rigorous theoretical framework for analysis of large-scale networks, ranging from the gene interaction network to the Internet [8-9]. Indeed, reconstructing and analyzing biological networks, be they of genes, proteins or cells, is at the heart of systems biology. The role of such "network biology" is to elucidate the processes by which complex behavior can arise in a system comprising mutually interacting components. While such *emergent* behavior at the systems level is not unique to biology [10], to explain properties of living systems, such as their robustness to environmental perturbations and evolutionary adaptability, as the outcome of the topological structure of the networks and the resulting dynamics, is a challenge of a different order. As networks appear at all scales in biology, from the intracellular to the ecological, one of the central questions is whether the same general principles of network function can apply to very different spatial and temporal scales in biology [11] (Fig. 1). In this article, we look at a few examples of how using a network approach to study systems at different scales can reveal surprising insights.

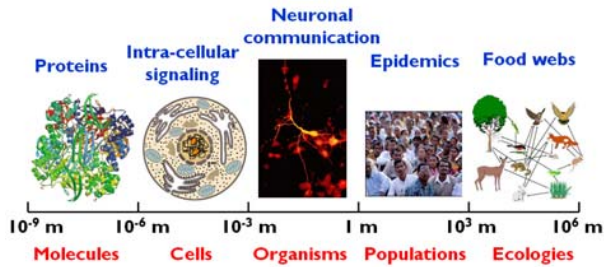

**Figure 1:** Dynamical processes mediated through complex networks occurring at different length scales in the biological world, from the protein contact networks at the molecular level to the network of trophic relations spanning several tens or hundreds of kilometers. In between these two extremes, we see intra-cellular networks (e.g., that involved in intra-cellular signaling), inter-cellular networks (e.g., neuronal networks in the brain) and networks among individuals comprising a population (e.g., the social contact network through which various epidemic diseases spread).

## INTRACELLULAR NETWORKS

Inside the biological cell, a variety of networks control the multitude of dynamical processes responsible for its proper function [12]. These include *genetic regulatory networks* [13,14] by which genes regulate the expression of other genes through activation or inhibition (i.e., by expressing proteins that act as promoters or suppressors of other genes) as seen, for example, in the pattern formation steps that occur during development from a fertilized cell into an embryo [15], *protein-protein interaction networks* involving the physical association of protein molecules that are vital for many biological processes such as signal transduction [16], and, *metabolic networks* of bio-chemical reactions [17-19], that are responsible for breaking down organic compounds to extract energy as well as those which use energy to construct vital components of the cell such as amino acids and nucleic acids. While the glycolytic pathway that converts glucose into pyruvate, the first significant portion of the metabolic network to be reconstructed, took many years to be elucidated, there are now experimental techniques such as the yeast two-hybrid screening method that test for physical interactions between many pairs of proteins at a time, allowing rapid reconstruction of such networks.

One of the most intriguing cellular networks is the protein-protein interaction network that is responsible for intracellular signaling, the mechanism by which a cell responds to various stimuli through an ordered sequence of biochemical reactions. These reactions regulate processes vital to the development and survival of the organism (e.g., differentiation, cell division, apoptosis, etc) by transmitting information from receptors located at the cell surface (that receive external signals) to specific intracellular targets in a series of enzyme-substrate reaction steps [20].

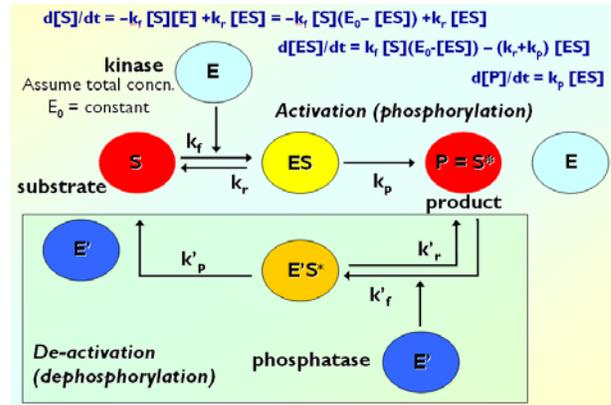

**Figure 2:** Dynamics of kinase activation (phosphorylation) and de-activation (dephosphorylation). The substrate kinase (S) is activated by its corresponding enzyme (E), which is the kinase located immediately upstream in the cascade. The enzyme-substrate complex (ES) formation is a reversible reaction with forward and reverse rates of $k_f$ and $k_r$, respectively, The product (P) of the enzyme-substrate reaction is the phosphorylated kinase S*, which is generated by an irreversible reaction step from ES with the rate $k_p$. The deactivation reaction from S* to S is mediated by the phosphatase E'. The enzyme-substrate reaction dynamics can be represented by the rate equations for the variation of concentrations of S, ES and P as shown at the top of the figure.

This dynamics is governed by a large number of enzyme molecules, which act as the nodes of the signaling network. Kinases belong to the most commonly observed class of enzymes involved in cell signaling and they activate target molecules (usually proteins, e.g., other types of kinase molecules) by transferring phosphate groups from energy donor molecules (e.g., ATP) to the target, the process being termed phosphorylation. The subsequent deactivation of the target molecules by dephosphorylation is mediated by corresponding phosphatases (Fig. 2).

In the post-genomic era, the focus of researchers has shifted from studying such molecules in isolation, to understanding how the set of interactions over the complex bio-molecular network can explain the entire repertoire of cell behavior [21]. An essential tool towards such an understanding is modeling the dynamics of reactions occurring in smaller subnetworks (*modules*) that act as distinctly identifiable functional units performing specific tasks [22]. The detailed knowledge of how individual modules control signaling dynamics can help us in eventually building a coherent picture for the functioning of the entire cellular network. Here, we focus on such a module, the three component mitogen-activated protein kinase (MAPK) cascade (Fig. 3), which is seen in all eukaryotic cells and is involved in many functions, including cell cycle control, stress response, etc. [23].

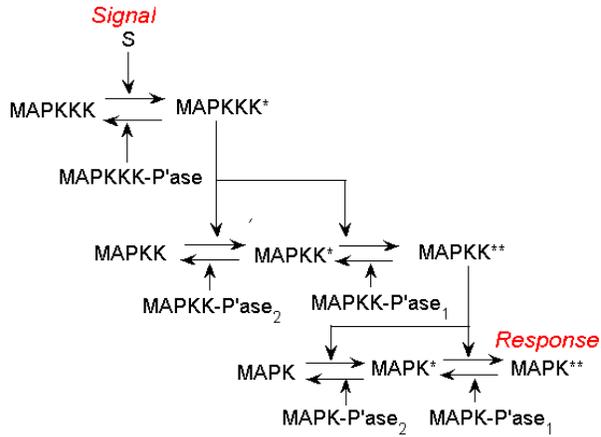
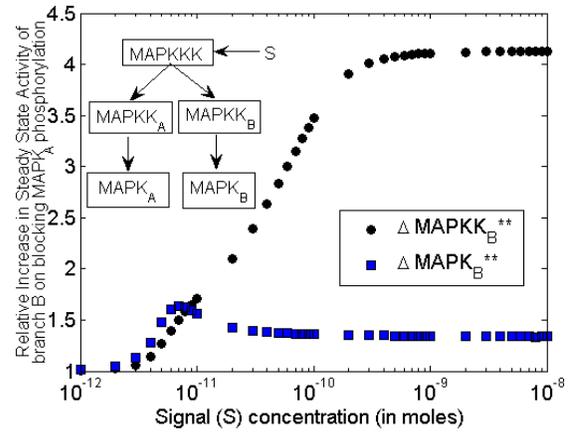

**Figure 3:** Schematic view of the MAP-kinase signaling pathway comprising MAPKKK, MAPKK and MAPK. The initial activation of the cascade is due to the signal S which initiates phosphorylation of MAPKKK, and the corresponding response is measured in terms of the degree of resultant activation in MAPK. The phosphorylated products are denoted by *, while P'ase indicates the corresponding phosphatases. Note the dual phosphorylation of MAPKK & MAPK, unlike the single phosphorylation of MAPKKK.

The single chain cascade, where an input (or signal) to the MAPK kinase kinase (MAPKKK) is transmitted via MAPK kinase (MAPKK) to the MAPK whose output is the activation of transcription factors or other kinases, is well understood. However, in many situations, e.g., in the network involved in processing stimulus received by the B-cell antigen receptor (a key player in human immune response), a branched variation of the basic MAPK module is observed. Here, a common MAPKKK sends signals to two different types of MAPKK and hence, to different MAPK molecules (Fig. 4, inset). We have analyzed a model of this branched structure to investigate how the elements in the two branches affect each other, although there is no direct interaction between them. We find that a novel mechanism of indirect regulation arises solely from enzyme reaction dynamics on the branched structure [24].

The time-evolution of the concentrations for different molecules in the branched MAPK network is described by a set of coupled ordinary differential equations (ODEs), with each enzyme-substrate (ES) reaction having a reversible ES complex formation step and an irreversible step of product formation (Fig. 2). The 32 coupled ODEs in our model are explicitly integrated numerically without invoking the quasi-steady-state hypothesis underlying Michaelis-Menten kinetics. The model parameters, viz., the different reaction rates and initial concentrations of the substrate molecules, are adapted from the Huang-Ferrell model [25]. MAPKKK is activated by a signal S whose concentration is varied in the physiologically plausible range of $10^{-12}$ to $10^{-8}$ moles.

**Figure 4:** The relative increase in the response of branch B as a function of signal strength, when the phosphorylation of MAPK in branch A is inhibited by an ATP blocker. The increase is measured as the ratio of steady-state concentrations for $MAPKK_B$** and $MAPK_B$** on inhibiting branch A to their concentrations under normal conditions. The inset shows a schematic diagram of the branched MAPK reaction network.

The system is first observed under normal conditions and then perturbed by preventing the activation of $MAPK_A$. As a result of blocking the phosphorylation of MAPK in branch A (through preventing the formation of $MAPK_A$* product from the $MAPK_A$-$MAPKK_A$** complex, which can be implemented in an experimental situation by using an ATP blocker), the concentration of free $MAPKK_A$ is significantly reduced. This is because, although $MAPK_A$ is not being activated, free $MAPK_A$ and its corresponding kinase $MAPKK_A$** are being steadily depleted from the system through ES complex formation. In the absence of product formation, the release of $MAPKK_A$** can only occur at the relatively slow rate of the ES-complex unbinding process. As increasing numbers of MAPKK** molecules are taken up into complexes with MAPK, there is less available free MAPKK for MAPKKK* to bind with. As a result, there is more MAPKKK* available for binding to $MAPKK_B$. This results in significant amplification of the activity in B branch, with all its downstream reactions being upregulated, including an increase in the dual phosphorylation of both $MAPKK_B$ and $MAPK_B$ (Fig. 4). This indirect regulation of activity between the two branches implies backward propagation of information along the signaling pathway from $MAPK_A$ to MAPKKK, in contrast to the normal forward direction of information flow from MAPKKK to MAPK.

Our results show that, in intracellular signaling networks there may be indirect regulation of activity between molecules in branched structures. Such long-range effects assume importance in light of the current paradigm in reconstructing intracellular networks where, the observation of up- or down-regulation of activity for a molecule as a result of perturbing another molecule is thought to be indicative of the existence of a direct interaction between them. As signaling networks are often inferred on the basis of such observations, our results provide an indication of other effects that can explain such dynamical correlations between the activities of different molecules.

**SYSTEMS NEUROSCIENCE**

Going up the scale from cellular to multi-cellular systems, we come across the systems biology questions related to inter-cellular communication and how such systems respond to events in the external environment. Possibly, the most intriguing questions in this domain have to do with the brain and the nervous system, the area of *systems neuroscience*. This field explores the neural basis of cognition, as well as of motivational, sensory, and motor processes. Systems neuroscience fills the gap between molecular & cellular approaches to the study of brain and the behavioral analysis of high-level mental functions. This definition is, however, nebulous as an enormous number of questions can be considered to be under the umbrella of systems neuroscience. E.g., such questions may concern issues of how to identify the neuronal correlates of consciousness and how multiple sensory stimuli define perceptions of reality, to much more tangible issues of the projection field of neurons. In an effort to classify the large variety of problems, Sejnowski and van Hemmen in a recent book [26] have divided them into questions about (i) the evolution of the brain, (ii) the organization of the cortex, (iii) computational ability of the brain, and (iv) the organization of cognitive systems. However, these classes cannot always neatly pigeonhole all systems-related questions about the brain, e.g., how do the computations taking place in the brain get translated into actions and decisions at the level of the individual organism.

Given the complexity, inaccessibility, and heterogeneity of the brain, systems neuroscience uses tools which are interdisciplinary in nature. While common behavioral experiments (e.g., involving the water-maze test for memory) and neuropsychological tests (e.g., showing sensory cues for studying synaesthesia) still form the basis for many investigations, technological improvements are allowing freedom in what and how these questions can be addressed. For example, new molecular and biophysical tools for the observation and manipulation of neural circuits allow one to image subjects as they are performing specific tasks. The field uses a variety of imaging techniques (fMRI, sMRI, DTI, and EEG) as well as behavioral, psychological and computational methods. The aim is to use all these tools to have a better understanding of the integrated functioning of large-scale distributed brain networks and how disruptions in brain function & connectivity impact behavior. Thus, systems neuroscience addresses questions on both normal and abnormal functioning of the nervous system, and also has the potential to significantly influence developments in the artificial intelligence community.

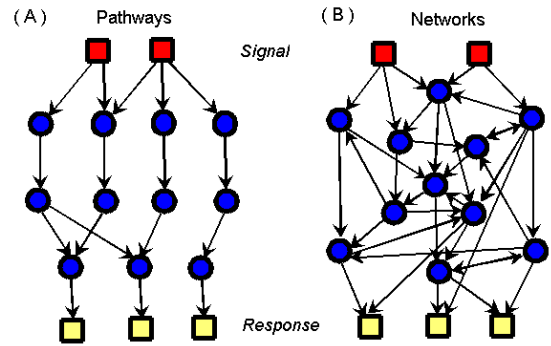

**Figure 5.** Biological information processing systems can adopt one of two possible structural arrangements: (A) a series of parallel pathways with relatively little communication (cross-talk) between them, and (B) a densely connected network. The squares indicate the input and output nodes of the system, while the circles indicate the intermediate nodes involved in signal processing.

Exploring cognitive ability and organization of the nervous system is obviously easier to study in a system with limited number of neurons. Not only does it allow the advantage of working with a numerically simpler system, but it also indicates what is the least number of neurons required for performing simple behavioral functions. The nervous system of the nematode *C. elegans* which has only 302 neurons, allows for tractability, unique identification of neurons and availability of a detailed physical connectivity map derived from ultrastructural analysis with electron microscope [27]. Computational tools can be used on such data for identifying patterns in the wiring of the C.elegans nervous system [28] as well as mapping the functional connectivity in specific behavioral circuits and delineating causal relationships between neural activity and behavior. This allows us to ask questions such as why does nature prefer to use networks rather than a parallel set of dedicated stimulus-response pathways to process information (Fig. 5). This is an example of how using the systems methodology allows one to see common themes in biological networks across length scales, as simply by changing the identity of the nodes one can ask the same question for the nervous system (nodes = neurons, links = excitatory and inhibitory synapses or gap junctions, stimulus-response path = reflex arc) and for the intra-cellular signaling network considered earlier (nodes = kinases, links = phosphorylation and dephosphorylation reactions, stimulus-response path = reaction pathway).

The core neural circuits behind some simple behaviors in the nematode have already been experimentally determined by selective and systematic ablation [29]. On the basis of these functional circuit neurons, we can aim to identify all neurons involved in a circuit for a particular behavior, the inputs that drive activity in each neuron, how input signals received in the sensory neurons are transformed as they travel through the circuit and finally how the pattern of neuronal depolarizations and hyperpolarizations over the entire network translates into behavior.

The functional circuits are seen to function in tandem rather than as isolated modules. To be specific, if one considers the eight functional circuits for (i) mechanosensation (touch sensitivity), (ii) chemosensation, (iii) thermotaxis, (iv) egg laying, (v) defecation, and, three types of locomotion, viz., when (vi) satiated (feeding), (vii) hungry (exploration) and (vii) during escape behavior (tap withdrawal), there are certain neurons which belong to more than one circuit. The interneurons commonly receive inputs from many modalities and are often multifunctional. However, the sensory neurons and sometimes even motorneurons may be dedicated for a particular function, as in the egg-laying circuit (occurring only in the hermaphrodite animal) which possess specialized motor neurons [27,30].

In a recent work [31], we have performed an integrated analysis of these functional circuits to discover emergent patterns in the connectivity profile of the C. elegans neural network. Using network core decomposition tools we observe correlations between a neuron having a critical functional role and it occupying a central position in terms of network structure. This is evidence for a structural basis of the roles played by neurons in the functional circuits. In order to identify the factors which determine the connectivity constraints in a neuronal network, we have looked at the modularity of the network with reference to position, function and neuron type. It appears that modules defined in terms of network connectivity do not necessarily correspond to the ganglia defined in terms of spatial location of the cell bodies. Thus, wiring economy and developmental constraints do not completely decide the connection structure of the network. However circuits sharing a large proportion of core neurons do show similarity in terms of their modular spectra which may be interpreted as a requirement to get input from these overlapping circuits in any perceptual decision-making process.

Our work also showed that while the network has a preferred direction of information flow, it is not a simple unidirectional propagation from sensory to motor neurons. The large number of recurrent connections observed among interneurons suggests a hierarchical structure with a densely connected core comprising mostly interneurons and less densely connected periphery populated by sensory and motor neurons. The betweenness centrality of the neurons increase with their core order membership, indicating that most of the shortest paths between pairs of neurons pass through neurons belonging to the innermost core. This would allow most of the information propagation to take place via a small select set which are subject to a high level of feedback activity. Such a hierarchical network (having a densely connected core and am overall sparse structure) also prevents indiscriminate global activation of the nervous system while at the same time permitting high density of connections to allow for high communication efficiency, so that information can propagate rapidly from sensory stimulation to motor reaction.

Based on anatomical, physiological and behavioral data, the simulation of the dynamics of neural circuits has already been attempted in a few organisms. In C.elegans, a dynamic network simulation of the nematode tap withdrawal circuit was worked out by Niebur and Erdos [32] while Ferree and Lockery [33] demonstrated that simple computational rules can predict certain components of chemotactic behavior. In spite of the ability to process a large variety of sensory inputs, the ultimate outputs of the C.elegans nervous system are simple. The organism performs a limited set of motor programs on integration of sensory information and there should be no indiscriminate activation of either sensory or interneurons of unrelated circuits. The challenge in front of systems neuroscience is to explicate the strategies used by networks to achieve this goal. The results of this investigation will be useful not only in the context of the nervous system but also for intra-cellular information processing networks such as those of kinases.